# Atomistic picture of fluorescent probes with hydrocarbon tails in lipid bilayer membranes: an investigation of selective affinities and fluorescent anisotropies in different environmental phases

S. Knippenberg[1,2,*], G. Fabre[3], S. Osella[4], F. Di Meo[5], M. Paloncýová[1], M. Ameloot[2], P. Trouillas[5,6]

[1] Division of Theoretical Chemistry and Biology, KTH Royal Institute of Technology Roslagstullsbacken 15, S-106 91 Stockholm, Sweden

[2] Biomedical Research Institute, Hasselt University, Agoralaan Building C, 3590, Diepenbeek, Belgium

[3] LCSN-EA1069, Faculty of Pharmacy, Limoges University, 2, rue du Dr. Marcland, 87025 Limoges Cedex, France

[4] Centre of New Technologies, University of Warsaw, Banacha 2C, 02-097 Warsaw, Poland

[5] INSERM UMR 1248, Faculty of Pharmacy, Limoges University, 2 rue du Docteur Marcland, 87025 Limoges Cedex, France

[6] Centre of Advanced Technologies and Materials, Faculty of Science, Palacký University, tř. 17 listopadu 12, 771 46 Olomouc, Czech Republic

(*) Corresponding author: sknippen@kth.se

**Abstract**

By reverting to spectroscopy, changes in the biological environment of a fluorescent probe can be monitored and the presence of various phases of the surrounding lipid bilayer membranes can be detected. However, it is currently not always clear in which phase the probe resides. The well-known orange 1,1′-dioctadecyl-3,3,3′,3′-tetramethylindodicarbo-cyanine perchlorate (DiI-C18(5)) fluorophore for instance as well as the new, blue BODIPY (4,4-difluoro-4-bora-3a,4a-diaza-s-indacene) derivative were experimentally seen to target and highlight identical parts of giant unilamellar vesicles of various compositions, comprising mixtures of dipalmitoylphosphatidylcholine (DPPC), dioleoylphosphatidylcholine (DOPC), sphingomyelin (SM) and cholesterol (Chol). However, it was not clear which of the coexisting membrane phases were visualized (Bacalum *et al.*, Langmuir 32 (2016), 3495). The present study addresses this issue by utilizing large-scale molecular dynamics simulations and the z-constraint method, which allows evaluating Gibbs free energy profiles. The current calculations give an indication why, at room temperature, both BODIPY and DiI-C18(5) probes prefer the gel ($S_o$) phase in DOPC/DPPC (2:3 molar ratio) and the liquid ordered ($L_o$) phase in DOPC/

SM/Chol (1:2:1 molar ratio) mixtures. This study highlights the important differences in orientation and location and therefore in efficiency between the probes when they are used in fluorescence microscopy to screen various lipid bilayer membrane phases. Dependent on the lipid composition, the angle between the transition state dipole moments of both probes and the normal to the membrane are found to deviate clearly from 90°. It is seen that the DiI-C18(5) probe is located in the headgroup region of the SM:Chol mixture, in close contact with water molecules. A fluorescence anisotropy study indicates also that DiI-C18(5) gives rise to a distinctive behavior in the SM:Chol membrane compared to the other considered membranes. The latter behavior has not been seen for the studied BODIPY probe, which is located deeper in the membrane.

**Introduction**

Molecular insight into the condition and properties of lipid membranes, which are fundamental components of living cells, is of utmost importance for various areas of biomedical research including drug design, drug pharmacology, or medical diagnosis and prognosis [1–4]. To give only one example of the crucial role of membranes, it has been shown that increased fluidity and polarity of cell membranes correlate with the metastasis in cancer cells [5]. Well-designed membrane-specific probes can picture biological membrane properties by means of optical imaging and suited spectroscopic techniques. Cholesterol (Chol) highly contributes to the structure of the membranes of many mammal cells [6]. For example, in hepatocellular carcinoma, which is the fifth most frequent cancer worldwide, high Chol levels were found to lead to tumor progression and malignancy [7–9]. The specific development of probe molecules which have an expressed affinity for Chol-abundant membrane regions is a particularly relevant and challenging topic. In the current work, computer modeling is used to investigate interactions of optically active probes with various membrane models and to evaluate whether they can identify the spectral fingerprints of specific biological conditions.

Natural membranes can be organized in different phases, with distinction between single-component and multicomponent membranes. For lipid systems of a single type, a gel phase ($S_o$) membrane is characterized by a high order of lipid packing. The liquid-crystalline or liquid disordered phase ($L_d$) of the membrane is characterized by a reduced lipid packing and higher diffusion coefficients. In complex lipid systems, Chol promotes phase segregation and gives access to the liquid-ordered state ($L_o$), a phase which is often also enriched in sphingomyelin (SM) [10–13]. The different ratios among the lipid components of a membrane are important parameters that determine its phase. Single component membranes made of dioleoylphosphatidylcholine (DOPC), dipalmitoylphosphatidylcholine (DPPC) or distearoylphosphatidylcholine (DSPC) have been extensively studied [12,14,15]. The phase in a single component system depends on the lipid chemical structure and the temperature. DOPC with its transition temperature of -17 °C adopts a liquid phase at room temperature; the DPPC membrane is in the $S_o$ phase at room temperature but adopts the $L_d$ phase above its transition temperature of 41 °C [16]. Two and/or three components membranes have also been evaluated, *e.g.*, made of DOPC/DPPC or DOPC/SM/Chol in studies, which highlighted that different ratios between the components modulate membrane properties [10,17–21]. Considering the number of possible combinations of membrane components, as well as possible lipid segregations, mixtures of phases are expected in biological membranes. The ternary mixtures can be schematically visualized along with their relevant tie lines in temperature dependent triangular diagrams, from which the phase compositions as well as their coexistence can be read [22–24]. There is finally the need of techniques capable of distinguishing and (locally) characterizing these different phases.

One of the most popular dyes to unravel this complex membrane structure is 1,1'-dioctadecyl-3,3,3',3'-tetramethylindodicarbocyanine perchlorate (DiI-C18(5)). It is a dialkyl carbocyanine (see Figure 1), which is amphiphilic due to the positively-charged head chromophore consisting of two indole rings connected by 5-carbon cyanine moiety and the 18-carbon saturated alkyl chains, which are important for the phase-selective partitioning in the membrane [25,26]. DiI-C18(5) exhibits a high extinction coefficient and a high fluorescence quantum yield, and is highly fluorescent and photostable when incorporated into membranes [27]. Fluorescence spectroscopic analyses of the DiI-family have been used to investigate: membrane rotational lipid mobility [28]; membrane potential [27]; membrane fusion [29]; fluorescence resonance energy transfer [30]; phase separation [31]; lipid leaflet transmigration [32]; and the existence of lipid rafts [33]. As the precise location, orientation and lipid/phase selectivity of the dye is often unknown or only partially described, the interpretation of fluorescence lifetime, anisotropy and rotational dynamics may be complex. Gullapalli *et al.* theoretically investigated the properties of two and four DiI-C18(3) probes, which have a cyanine backbone made of 3 carbon atoms, within a DPPC lipid bilayer in its $L_d$ phase at 323 K, safely above the transition temperature [34]. The probes were found below the head group – water interface and report well the rotational and lateral diffusion components of the lipid dynamics. The calculations showed that the dye causes minor changes at the interface in the ordering of the water dipoles and electrostatic potential.

Recently, the *meso*-amino substituted BODIPY probe 8-[(2-sulfonatoethyl)amino]-4,4-difluoro-3,5-dioctadecyl-4-bora-3*a*,4*a*-diaza-*s*-indacene (BNP, see Figure 1) was synthetized and optically characterized [35]. This probe expresses similar behavior with respect to membranes as DiI-C18(5), but fluoresces in the blue part of the visible spectrum. The BODIPY dyes are known to combine outstanding spectroscopic and (photo)-physical properties, such as bright fluorescence with absorption and emission bands in the visible range, as well as stability toward light and chemicals. In particular, BNP was found to be excitable by either 1 or 2 photons in combination with a high fluorescence quantum yield; this probe was found to preferentially partition in the same lipid phase as DiI-C18(5) [35]. In this experimental work by Bacalum *et al.*, BNP and DiI-C18(5) were studied in a 2:3 mixture of DOPC:DPPC ($L_d$:$S_o$ phases) and a 1:2:1 mixture of DOPC:SM:Chol ($L_d$:$L_o$ phases) at room temperature. Although we could expect a tiny contribution of DOPC to the $L_o$ phase, for simplicity it has been further omitted. Li and Cheng observed that the smaller DiI-C18(3) probe preferentially partitioned in the DPPC $S_o$ phase of the DOPC:DPPC binary mixture [36]. With respect to the ternary mixture, Baumgart *et al.* investigated a 50:27:23 ratio (DOPC:SM:Chol), and reported that DiI-C18(3) preferentially partitions in the DOPC $L_d$ phase [17]. Fluorescence microscopy provided insights into the DiI-C18(3) probe embedded in a dozen ternary mixtures [18]. However, neither for the larger DiI-C18(5) probe nor for BNP, the phase partitioning is known for the specific ratio of lipid systems considered in [35].

It is currently a challenge to accurately evaluate optical properties of the probe within various lipid bilayers. This task first requires a correct and comprehensive evaluation of large scale structural features of the molecular assembly made of the probe and the lipid bilayer, which can be obtained by Molecular Dynamics (MD) simulations. For the current work therefore, MD simulations were performed to gain insight into the interactions of specifically both DiI-C18(5) and BNP probes within biological membranes and to understand their phase preference. Attention is paid to their locations and motions within the lipid bilayers and how this impacts on their spectroscopic features. *In silico* membrane models have been constructed in the past and a vast development with increasing accuracy is noted [37–42]. MD calculations have been used to accurately evaluate simultaneously equilibrium positions of xenobiotics in lipid bilayers, their partition and diffusion coefficients at subpicosecond and atomic resolution [43–

[48]. Focusing on the simulation of optically active probes opens the possibility towards the development of non-invasive techniques which provide insights into the impact of surrounding environment in (non) linear and fluorescence spectroscopy [49,50]. Here, MD simulations are used to assess the interaction of both DiI-C18(5) and BNP in four different lipid bilayers and lipid phases. One of them is the DPPC membrane in its Ld phase, which is considered at the same temperature as in the study of Gullapalli *et al.* [34] to enable a direct comparison. The structural and physical-chemical properties of the four lipid bilayer models are discussed in terms of their areas per lipid, order parameters and non-bonding interaction energies. The Gibbs free energy profiles of DiI-C18(5) and BNP are investigated along the *z*-axis of the membrane, which is oriented perpendicular to the membrane surface. The differences between the equilibrium positions and orientations of both probes, and the variations of their transition dipole moments within the various environments are identified as being decisive for the linear and non-linear optical spectra [51,52]. Finally, the fluorescence anisotropy of both probes is modelled and similarities as well as differences in the behavior of DiI-C18(5) and BNP are highlighted.

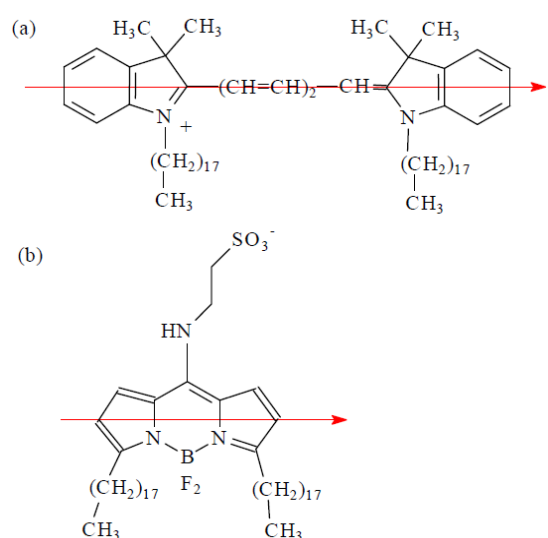

**Figure 1: Molecular structures of (a) DiI-C18(5) and (b) BNP. The red arrows are the transition state dipole moments for both probes. To describe the positions of the probes, the middle carbon atom of the –(CH=CH)$_2$–CH= bridge is considered for DiI-C18(5) and the Boron atom for BNP. Remark that the π-conjugated core in both molecules is confined to those parts of the molecules without tails and – in the case of BNP – without headgroup.**

**Computational details**

The MD simulations were performed using the Gromacs 4.5.7 [53,54] software and the Gromos 43A1-S3 force field [55–58]. The lipid bilayer models consisted of 128 lipid molecules surrounded by at least 4500 explicit water molecules, which were described by the extended single point charge (SPC/E) model. $Na^+$ and $Cl^-$ ions were added to bulk water at a physiological concentration (0.9%). The spatial reference frame is such that the *x*- and *y*-axes are taken in the plane of the bilayer, whereas the *z*-axis is perpendicular to the membrane surface. Periodic boundary conditions were considered in 3 dimensions. Electrostatic interactions were treated by the particle-Mesh Ewald method [59] and bonds were constrained by the LINCS algorithm [60]. Electrostatics and van der Waals short-range interaction cutoffs were set to 1.6 nm. The NPT ensemble was used, with the Nosé−Hoover thermostat [61,62], and a Parrinello−Rahman barostat [63] for a semi-isotropic pressure coupling at 1 bar and compressibility of $4.5 \times 10^{-5}$ $bar^{-1}$. The simulation time step was set to 2 fs and the coordinates in the simulation were saved every 500 steps.

The four lipid bilayer models were built with a homemade script, they consisted out of one probe and in total 128 lipids (two leaflets of 64 lipids): pure DOPC at 298 K ($L_d$ phase), pure DPPC at 298 K ($S_o$ phase) and at 323 K ($L_d$ phase), and a 2:1 SM:Chol mixture at 298 K ($L_o$ phase). The SM acyl chains contain 17 and 15 methyl groups for the sn1 and sn2 acyl chains, respectively. Upon these systems, periodic boundary conditions in all directions have been applied. All membranes were equilibrated during 20 to 40 ns long free simulations, after which convergence of structural parameters (*i.e.*, area per lipid, lipid order parameters…) were ensured. In line with previous work [35], atom types were assigned by PRODRG [64], while partial atomic charges have been used which result from the restrained fit of electrostatic potential (RESP) [65]. They were calculated at the level of density functional theory (DFT) by means of the B3LYP functional [66,67], Dunning's correlation consistent cc-pVDZ basis set [68], and a PCM model which was chosen to describe an implicit solvent model with a dielectric constant of diethyl ether ($\varepsilon = 4.24$) [69]. The Lennard-Jones parameters of the boron atom for the BNP probe, which are not by default present in the applied force field, have been taken from reference [70]. Further parametrizations for the bonded interactions of the Boron atom have been performed by means of previous DFT method.

The Gibbs free energy profiles for BNP and DiI-C18(5) were calculated by means of the *z*-constraint method [71,72], in which bulk water was put as a reference. The distance between the centers of masses of the lipid bilayer and the Boron atom for BNP, or the middle carbon atom of the –(CH=CH)2–CH= bridge for DiI, was constrained, and the required force was monitored.

The averaged force was then used to calculate the Gibbs free energy profile, also called potential of the mean force [72,73], as:

$$\Delta G(z) = - \int_{outside}^{z} \langle F(z')\rangle_t \, dz',  \qquad (1)$$

where $\langle F(z)\rangle_t$ is the force which is needed to keep the molecule at a given depth $z$. A series of windows was obtained every 0.1 nm for $z$-constraint simulations. The initial structures for each window were generated by merging probe and membrane coordinates, minimized to avoid steric clashes, and *g_membed* [74] was used to remove overlapping lipids when appropriate. In this process, the probes were oriented along the $z$-axis, with the lipid tails in the direction of membrane center. For the $z$-constraint process, 100 ns simulations were performed per window, ensuring convergence of Gibbs free energy profiles. The computational error was found to be ~1 kcal/mol. Starting from the minimum energy positions of the Gibbs free energy profiles, 300 ns long MD simulations were performed without applying additional constraints, of which the first 40 ns were discarded from the simulation window, as being the time required to equilibrate the system. The analysis of the structures of the membranes were performed on these unbiased simulations with GROMACS internal tools, area per lipid for individual lipid types was obtained by the FATSLiM script [75].

The transition state dipole moments of the BNP and DiI-C18(5) probes have been calculated using approximate second order coupled cluster theory (CC2) and the double zeta polarized (DZP) basis set.

In total, these simulations required a computational effort of more than 40 μs. To perform these calculations, the *Lindgren* cluster at the PDC Center for High Performance Computing in Stockholm (864 000 core hours, 2013-2014), the *muk* tier-1 cluster of the Flemish Supercomputer Centre (VSC) (264 960 core hours, 2014-2015), as well as the *Beskow*, *Triolith* and *Abisko* clusters with in total 105 000 core hours/month (2015) were used.

**Results and discussion**

**Characterization of the membranes**

If the simulated DOPC ($L_d$), DPPC ($S_o$), DPPC ($L_d$) and SM:Chol ($L_o$) bilayer membranes are expected to influence the distribution and dynamic behavior of the embedded probes, their inherent properties should be accurately modelled. The structure of lipid membranes can be well described by the density plots of various membrane components along the normal axis to the membrane plane. The density distributions of the lipid constituents and of water from the center of the membrane were constant between 2 nm and 0.8 nm for SM:Chol (Figure 2). In the other three membranes, locally higher lipid density was found with a peak at around 1.7 nm from the center, followed by a rapid decrease to the center. This effect is explained by the presence of free volumes just beneath the aqueous interface in contact with the polar head group region [34,76,77] and is manifestly seen in the SM:Chol membrane. Concomitantly, the thickness of the SM:Chol membrane was greater, as seen by a shifted point where the density of the water equals that of the lipids (*i.e.*, crossing at 2.5 nm for SM:Chol with respect to 2.3 nm for the $L_d$-phase DOPC and DPPC, see Figure 2). As expected, a similar increase of the thickness was observed for DPPC ($S_o$).

The thickness, in terms of distance of the highest density peaks, agrees well with experimental data. We observed differences in thickness between the different membranes, namely 4 and 4.5 nm for the DPPC ($L_d$) and the SM:Chol bilayer, respectively. The latter simulated thickness agrees with the experimental value of 4.6-4.7 nm [78]. This value mainly depends on SM, as Chol is known not to significantly modify the conformation of SM molecules [14]. The thickness of the DOPC and DPPC ($S_o$) bilayers is found in between 4 and 4.5 nm.

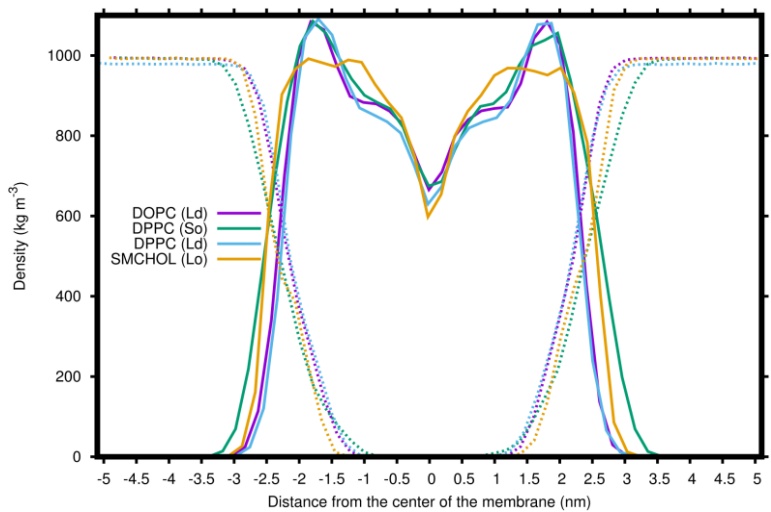

**Figure 2: Density distributions of the lipid constituents (full line) and of water (dotted line) within the various membrane phases with respect to the center of the membrane**

Over the 300 ns of MD simulations, the area per lipid exhibited constant values, *i.e.*, ~0.45 nm$^2$/lipid for SM:Chol ($L_o$, 0.40 nm$^2$/Chol and 0.48 nm$^2$/SM), ~0.51-0.52 nm$^2$/lipid for DPPC ($S_o$), 0.58 nm$^2$/lipid for DOPC and DPPC ($L_d$) (Figure S1). Although the calculated area per lipid in the $L_d$ phase is lower than some of the reported experimental data [79], ],the area/lipid values represent well the differences in the studied phases. A tighter packing and condensing effect were previously observed in $S_o$ phase as well as in the presence of Chol [80].

The potential energy of interaction between the lipid tails ($V_{tails}$) can be derived from the average sum of Lennard-Jones and short-range Coulomb potentials between all pairs of atoms in the lipid tail region [80,81]. Concerning Chol, all atoms but the hydroxyl group were included, whereas for phospholipids, all tail atoms up to the three glycerol carbon atoms were included. The potential energy was averaged from 180 to 280 ns. The $V_{tails}$ values per atom are very similar for all four membrane models (1.193, 1.127 and 1.110 kcal/mol for DPPC ($S_o$), SM:Chol and DPPC ($L_d$), respectively). The decreasing values from 1.193 to 1.110 kcal/mol mainly point the decrease of van der Waals contacts between lipid tails. The latter value is different from that of DOPC, which amounts to 1.182 kcal/mol, likely be due to the greater van der Waals interactions between the unsaturated bonds deep in the DOPC tails not present in the DPPC molecules. Finally, the value for DPPC ($S_o$) agrees with the one communicated by Wennberg *et al.* in 2012 [80].

To characterize the employed membrane models, the order parameter $|S_{CD}|$ is calculated, too. It is experimentally obtained using deuterium NMR by using the equation [34,82]

$$S_{CD} = \left\langle \frac{3}{2}(\cos^2 \theta_{CD}) - \frac{1}{2} \right\rangle, \qquad (2)$$

with $\theta_{CD}$ being the angle between C-H bond of the lipid tails and the *z*-axis. The brackets denote time averaging and corresponds to an ensemble averaging when experiments are performed. The value of the order parameter $S_{CD}$ can vary from -0.5 with $\theta_{CD} = 90°$ (indicating full ordering of the C-H bonds perpendicular to the z-axis and to a lesser extent an orientation of the C-C bonds along the *z*-axis) to 1 with $\theta_{CD} = 0°$ (indicating full ordering of the C-H bonds along the *z*-axis and the C-C bonds therefore more oriented perpendicular to the *z*-axis). Based on $S_{CD}$ values, we confirmed the typical differences between the membranes in the $L_d$, $S_o$ and $L_o$ phases: as reported in Figure S2, $|S_{CD}|$ values for the sn-1 and sn-2 tails amount maximally to ~0.40 for SM:Chol, ~0.35 for DPPC ($S_o$), and 0.25 for both DOPC and DPPC ($L_d$). These maxima are obtained at carbon C8 for SM:Chol and DPPC ($S_o$), while for DOPC and DPPC, the maxima are reached at C6. For C3, close to the headgroup and the glycerol moiety of the lipids, $S_{CD}$ amount to 0.27 for SM:Chol as well as for DPPC ($S_o$), and to 0.20 for both $L_d$ membranes. For SM:Chol, the quite strong increase in $|S_{CD}|$ towards the middle of the tail can be linked with the presence of Chol, which pushes the tails of SM deeper in the membrane, so as to accommodate the perpendicular orientation of the C-H bonds, diminishing hydrophobic effects. On the other hand, for DPPC ($S_o$), the high $|S_{CD}|$ values are related to the high packing, in agreement with $V_{tails}$ values, and with the higher amount of water present at the level of the glycerol group of the tails (Figure 2).

**Gibbs free energy profiles for DiI-C18(5)**

z-Dependent Gibbs free energy profiles provide information about partition and preferred positions (free-energy minima), as well as capacity of transfer from one to the other leaflet (Gibbs free energy barriers) independently from diffusion effects. The profile for DiI-C18(5) (Figure 3, left hand side) exhibits the deepest well (-38 kcal/mol) in the DPPC ($S_o$) membrane. The well is energetically less favorable by 5 kcal/mol in both the DPPC ($L_d$) and SM:Chol ($L_o$) bilayers; therefore based on the Gibbs free energy alone, one cannot distinguish any preferred affinity to both DPPC ($L_d$) and SM:Chol ($L_o$) bilayers. The affinity of DiI-C18(5) to DOPC ($L_d$) membrane is the least favorable one (potential well of -28 kcal/mol). The here presented data seem to answer therefore the question which membrane DiI-C18(5) prefers in a DOPC:DPPC ($L_d$:$S_o$) and a DOPC:SM:Chol ($L_d$:$L_o$) mixture, like has been used by Bacalum *et al.* in ref. [35]. Namely, the simulations indicate that in the former case, after equilibration of the biological environment, confocal microscopy will allow visualizing the DPPC ($S_o$) regions of the unilamellar vesicle, whereas in the latter case, the $L_o$ region of the SM:Chol mixture will be bright. For the concentrations used in the current study, DiI-C18(5) should thus be considered as a $L_o$ marker, and contrasts therefore with the findings of Baumgart *et al.* and Kahya *et al.* for DiI-C18(3) embedded in ternary lipid mixtures with other concentration ratios [17,18].

From the analysis given in Figure 3, the position of the global minima were similar except for SM:Chol (1.3, 1.3, 1.2 and 1.9 nm for DOPC ($L_d$), DPPC ($S_o$), DPPC ($L_d$) and SM:Chol ($L_o$), respectively). Although in this latter case, the bilayer thickness is greater, this makes DiI-C18(5) closer to the polar group region in SM:Chol with respect to the other membranes.

As we applied the *z*-constraint method from the center of the membrane and used a window for every Ångström, the barriers of transfer from one to the other leaflet have been obtained.

Significant differences are seen: the barrier at the middle of the bilayer is ~8 kcal/mol with DOPC and SM:Chol, and it is lower (4-5 kcal/mol) with DPPC ($L_d$) and DPPC ($S_o$). As repeatedly seen with amphiphilic compounds, the insertion into fluid bilayers requires small or even no energetic barriers in the polar head group region. Noncovalent interactions (electrostatic and H-bonding) mainly drive insertion and positioning, with little influence of size within the µs timescale.

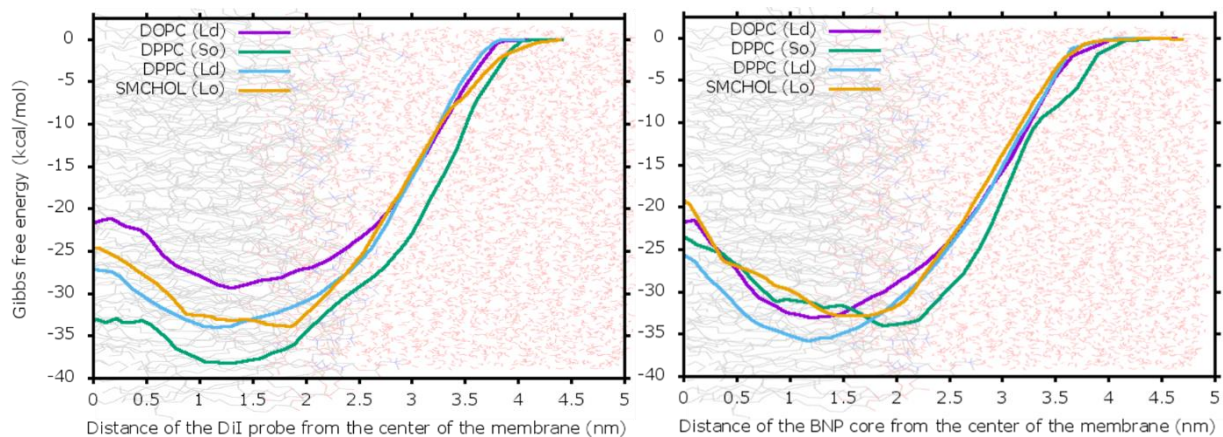

**Figure 3: Gibbs free energy surfaces of (left) DiI-$C_{18}$(5) and (right) BNP in function of the distance (in nm) from the center of the membrane along the *z*-axis, perpendicular to the membrane surface. The centers of mass of the DiI-C18(5) and BNP cores have been constrained. The error bar is contained in the thickness of the line.**

**Analysis of the unconstrained trajectories for DiI-C18(5) in the various membranes**

It is worth noting that the Gibbs free energy profiles are generated based upon a constrained movement of the core of the probe. To discuss the equilibrated positions and orientations of DiI-C18(5) and to profoundly evaluate the influence of the finite temperature, a free production run of 300 ns was performed for each membrane in the presence of DiI-C18(5), with the minima of the Gibbs free energy profile as starting geometries. Illustrations of the DiI-C18(5) probe in the various membranes are given in Figure 4. As a measure for the position of DiI-C18(5), the middle carbon atom of the cyanine-backbone was considered with respect to the membrane center. For SM:Chol, DiI-C18(5) is situated at 1.75±0.11 nm from the membrane center, in close contact to the polar head group region (Figure 5). For both DPPC bilayers, DiI-C18(5) is located deeper, at ~1.0 nm from the membrane center, *i.e.*, in contact with the lipid tails (the exact value for the $S_o$ is 1.04±0.09 nm, while it is 1.09±0.09 nm for the $L_d$ phase). Gullapalli *et al.* observed a value which was with its 1.26 nm a bit higher for DiI-C18(3) in DPPC ($S_o$) [34]. In DOPC, the location is an intermediate of the other two, however with a broad distribution ranging from 1.3 to 1.8 nm (1.47±0.21). Except for SM:Chol, the mean positions in free simulations were slightly deeper than the positions of the free energy minima, but these differences were found within errors and thermal motion. In all membranes, the probes have their light sensitive core embedded in lipid head groups and the lipophilic tails pointing towards the center of the membrane. We calculated the angles of the tails of DiI-C18(5) with the *z*-axis (Figure S3). These angles take the value ~155° for both $L_d$ membrane phases, ~165° for DPPC in the $S_o$ phase, and ~170° for SM:Chol in the $L_o$ phase.

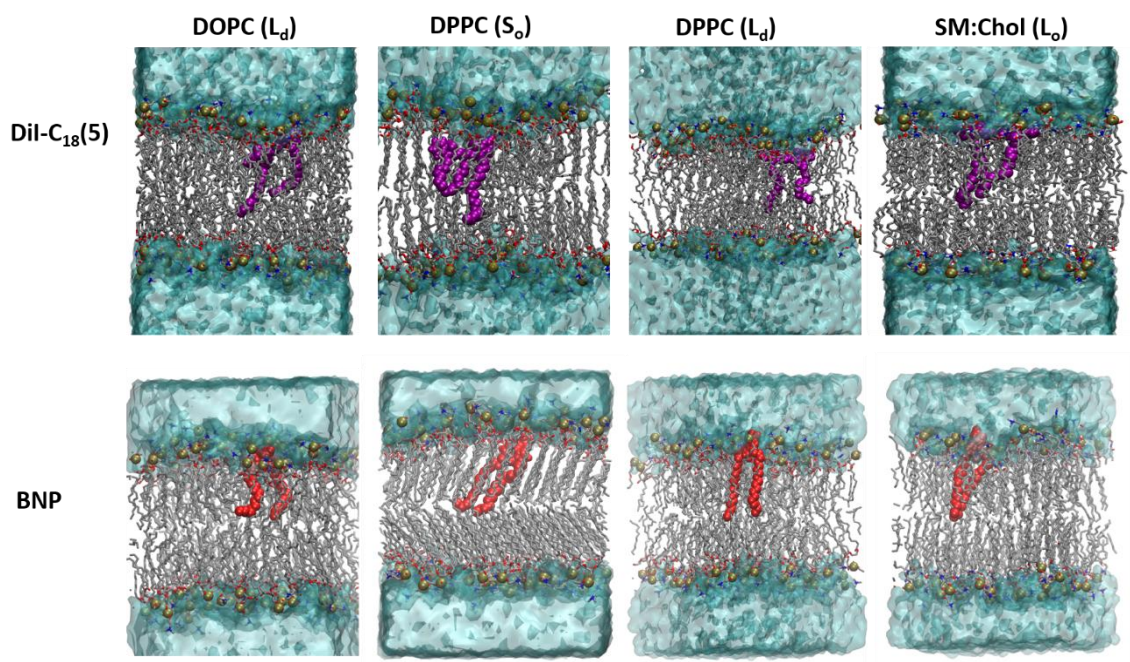

**Figure 4: Illustrations of the DiI-C18(5) and BNP probes in the different environments under investigation in the current study.**

While in SM:Chol the chromophore moiety of DiI-C18(5) is located at the surface of the membrane in contact with bulk water, it is located significantly more deeply in the other membranes. The DiI-C18(5) π-conjugated core is located below the level of the phosphates at a distance of 0.5 nm in SM:Chol, 0.3-0.8 nm in the $L_d$ phase of DOPC, 1.0 nm in the $L_d$ phase of DPPC and 1.2 nm in So phase (DPPC). In the SM:Chol membrane one has to consider not only the average level of membrane surface, but also a local arrangement of the membrane. The chromophore moiety of DiI-C18(5) experiences here free volumes and induces a small cavity, in which water molecules are pulled (Figure S4). Indeed, due to this surface position and such re-arrangements, DiI-C18(5) is more surrounded by water molecules in the SM:Chol membrane than *e.g.* in the $S_o$ phase or the $L_d$ phases of DPPC. For SM:Chol, at the DiI-C18(5) preferred position, the density of water is still 45% of that of the pure water layer, while practically no water is left with DPPC both in $S_o$ and $L_d$ phases (Figure 2). It can also be remarked that the maximum density of water experienced by DiI-C18(5) in the DOPC ($L_d$) membrane amounts to 20%. This effect is quantified by the radial distribution functions of DiI-C18(5) and the surrounding water molecules in the various membranes (Figure S5): the first maxima (at 0.45 nm from the DiI-C18(5) core) is very low for both phases of DPPC membranes (<0.2), slightly higher in DOPC (0.3) and significantly higher in SM:Chol (0.6). The water cavity experienced in the SM:Chol membrane is then responsible for the different behavior of DiI-C18(5) in this membrane.

Being decisive for the photoselection of the probe, the distribution of the angles between the transition dipole moment and the *z*-axis of the membrane is given in Figure 5. For DiI-C18(5), the transition dipole moment is oriented along the cyanine backbone and is displayed in Figure 1. Knowing that a perfect photoselection in confocal microscopy requires an angle of 90°, DiI-C18(5) in DOPC appears the most efficient with a most populated angle of ~85°. For SM:Chol and DPPC ($S_o$), the most abundant peak is seen at 72°. It can be remarked that for DPPC ($S_o$), the distribution of the angle is rather symmetric around its maximum, while for SM:Chol a slight asymmetry is seen together with a minor shoulder at higher values. The DPPC ($L_d$) lipid bilayer is characterized by a broad distribution of angles of a similar population, which are between 70° and 80°, which agrees with the angle of 77° reported for the smaller DiI compounds investigated by Gullapalli *et al.* [34] or with the range of ±10° around the perpendicular position with respect to the z-axis reported by Axelrod for erythrocyte ghosts [83]. The pronounced angles of the transition state dipole moments in the different membranes can be related to the differences in orientation between the sn-1 and sn-2 chains of the lipids and to the differences in position of the probe along the z-axis. To better describe the orientation of DiI-C18(5), the angle between the normal to the coplanar core and the *z*-axis was followed as well. A symmetric distribution was obtained centered at around 51° only for SM:Chol. In DOPC, essentially all values between 30° and 80° were observed, with only a slight preference for 35-40°. For DPPC ($L_d$), the angle increased from 30° to 80°. In DPPC ($S_o$), the angle distribution was ranging from 70° to 80°. Combining the analyses for both angles, $S_o$, and to a less extend $L_o$, restrain orientation to the probe.

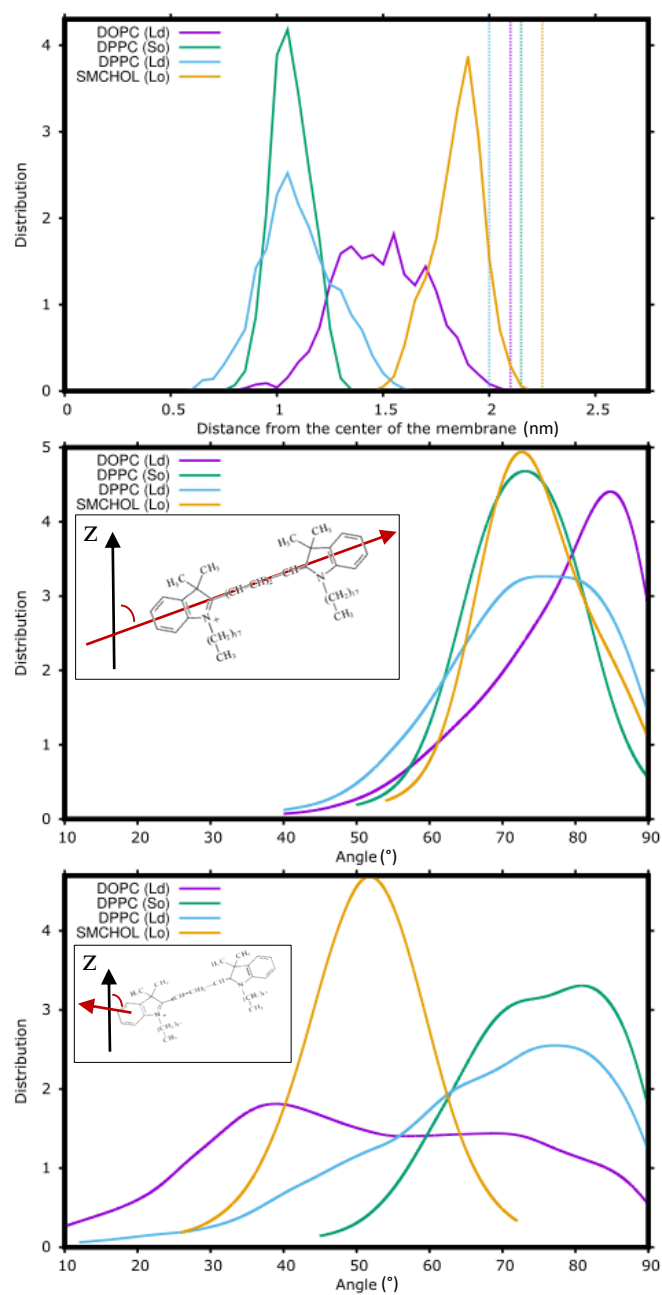

**Figure 5: DiI in various membrane phases** – (top) the position of the middle atom of the cyanine backbone of DiI along the *z*-axis expressed in terms of the distance from the center of the membrane, the dotted vertical lines denote the most abundant position of the phosphor atoms; (center) the angle between the transition dipole moment and the *z*-axis; (bottom) angle between the axis perpendicular to the plane of the DiI-C18(5) molecule and the *z*-axis. These data are taken from a free MD run and are convoluted with Gaussian profile peaks with a full width half maximum of 8°. The errors are displayed in Figure S6.

The order parameter profiles of DiI-C18(5) show the same trend in all four domains, *i.e.*, higher values close to the polar head group region which decrease when inserting deeper in the bilayer, as expected for lipid-type compounds (Figure 6). Close to the polar head, the highest $|S_{CD}|$ values (0.35-0.39) are observed in SM:Chol ($L_o$) and DPPC ($S_o$), whereas lower values (0.20-0.23) are observed in DOPC and DPPC ($L_d$). A further analysis can be performed making use of the above definition of $S_{CD}$ which relates to the angles between C-H bonds of the lipid tails and the $z$-axis. Due to the free space which is available at the top of the SM/Chol bilayer and the high abundance of water molecules, the mid C-C bonds of the tails of DiI-C18(5) are seen to straightly enter further down towards the center of the membrane, parallel to the z-axis. In DPPC ($S_o$), the DiI-C18(5) $|S_{CD}|$ value is also high for the first bonds below the nitrogen atoms, but the curve flattens down and the slope diminishes due to the high packing between the lipid tails, assuring a well-defined and orientation of the last carbon-carbon bonds of the tails. As expected from the position of the probe and the characteristics of the subsequent lipid bilayers, $|S_{CD}|$ values are lower in both $L_d$-phase membrane models, while the typical decrease along the tails is less steep than for the other two lipid bilayers. In the DOPC membrane, in the middle of the tails of DiI-C18(5), a slight increase of the $S_{CD}$ value is further on observed, which even surpasses the corresponding values for DiI-C18(5) in the DPPC ($S_o$) environment, which is in DOPC attributed to the double bond.

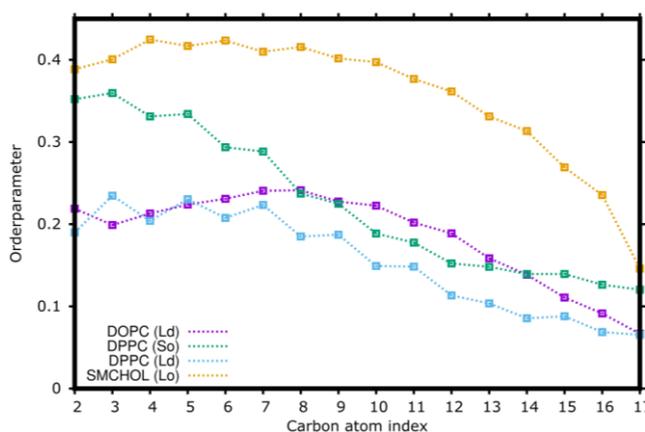

Figure 6: Order parameters for DiI-C18(5) in the various membranes. The carbon atom index points at the number of the carbon in one of the tails, starting from the carbons attached to each of the nitrogens.

**Gibbs free energy profiles for BNP**

The Gibbs free energy profiles of BNP given in at the right hand side in Figure 3 show a well of -33 kcal/mol in DOPC and SM:Chol; it is marginally deeper in DPPC ($S_o$) and significantly deeper in DPPC ($L_d$). Also, the most stable positions of BNP can to some extent be identified within the limits of the used $z$-constraint method. It mostly partitions at 1.2, 1.9, 1.1 and 1.5 nm in DOPC ($L_d$), DPPC ($S_o$), DPPC ($L_d$) and SM:Chol ($L_o$), respectively. The differences in the preferred position are however less clear than for DiI-C18(5). The markedly small Gibbs free energy differences in these profiles illustrate why within the constraints of the employed theories and simulations a comparison with DiI-C18(5) was needed to identify the lipid phases present in the bright areas which were seen in the confocal microscopy images published in [35]. Based upon our simulations for DiI-C18(5) and the related discussion above, it is subsequently safe to assume that BNP in the employed biological environments can be found in the $S_o$ phase when a mixture of DOPC ($L_d$) and DPPC ($S_o$) is considered and in the $L_o$ phase when a mixture of DOPC ($L_d$) and SM/Chol ($L_o$) is involved. We would like to stress here again the importance of the ratio of the employed mixture, as we employed DOPC:SM:Chol in a 1:2:1 ratio. By means of comparison, Baumgart *et al.* reported the $L_d$ phase as the preferred one for the DiI-C18(3) probe in a DOPC/SM/Chol mixture in basically a 2:1:1 ratio [17]. Other authors who reverted to the benchmark DiI-C18(5) probe, discussed the ternary mixtures in other ratios, too, without solving the issue for the mixture under investigation in the current study, but warning for the particular strong influence of the mixed lipid constituents when phase preferences are concerned [18–21].

The barrier for the transfer of BNP between the upper and lower leaflet amounts to ~10 kcal/mol for both DPPC membranes as well as for DOPC. It is calculated as the difference between the minimum of the potential energy surface and the maximum Gibbs free energy value found around the membrane center. The barrier amounts to ~14 kcal/mol for SM:Chol ($L_o$). The largest differences between both probes are therefore found for DPPC ($S_o$) and SM:Chol ($L_o$); the larger barriers are here reported for BNP and should be allocated to the influence of the Boron and Fluorine atoms.

**Analysis of the unconstrained trajectories for BNP in the various membranes**

As for DiI-C18(5), selecting the frames from the global minima of the Gibbs free energy profiles, a free production run was performed for 300 ns. Illustrations of the BNP probe in the various membranes are given in Figure 4. The boron atom of BNP was at 1.4-1.5 ± 0.2nm from the membrane center in both the DOPC and SM:Chol membranes (Figure 7), while it was inserted deeper (at 1.2 ± 0.2 nm) in DPPC ($L_d$). Conversely, in DPPC ($S_o$), it was at ~1.7 ± 0.1 nm, closer to the phosphorus atoms of the membrane surface, being located at 2.25 nm. It can be remarked that the boron atom is located rather close to the lipid tails, while the middle atom of the cyanine backbone of DiI-C18(5) is found higher in the molecule.

This difference in preferred position in the DPPC ($S_o$) and DPPC ($L_d$) environments is related to the difference in packing and area per lipid between both membranes. In DPPC ($S_o$), the packing in between lipid tails is likely to complicate insertion of BNP. Moreover, the core of BNP has a weak zwitterionic charge distribution between the nitrogen and boron atoms, making them slightly positive and negative, respectively. This favors interactions with water molecules abundant in this region of DPPC ($S_o$) (up to 20% of the density of pure water). The similar position of BNP in DOPC and SM:Chol is a manifestation of the interaction with tail unsaturation and Chol.

The angle between the transition dipole moment of BNP and the $z$-axis amounts to 70°-75° in DOPC (Figure 7). With an angle of 85° (and a minor distribution at 50°), the photoselection was found to be stronger in DPPC ($L_d$). In the DPPC ($S_o$) bilayer, the maximum of the distribution is found at 67°, however a shoulder can also be seen at 86°. Rather in contrast to DiI-C18(5), the angle distribution in SM:Chol is very broad with many contributions between 30° and 60°, and a major peak at 73°.

The orientation of the molecular plane of BNP with respect to the $z$-axis showed that this probe is rather perpendicular to the surface, with an angle of ~85° for DPPC ($S_o$) and SM:Chol. In DOPC, the maximum is at ~71°, although a shoulder is noticed at 85°. In DPPC ($L_d$), the distribution is broader, with a shallow maximum at 59°.

Although it has been experimentally found that both DiI-C18(5) and BNP probes target the same membrane phases and in contradiction to the first assumptions [35], it can be concluded based upon the current MD simulations that BNP behaves rather differently from the relatively known DiI-C18(5) one in terms of its orientation and equilibrium position in the membrane.

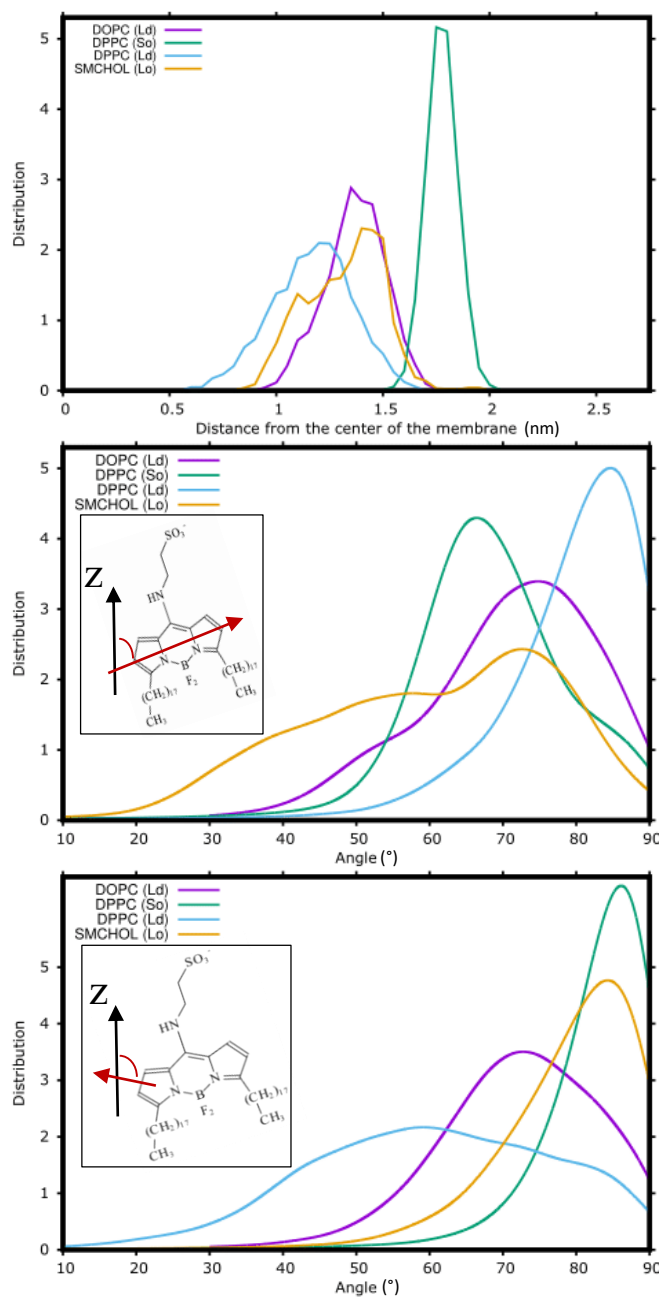

**Figure 7: BNP in various membrane phases –** (top) the position of the boron atom along the *z*-axis expressed in terms of the distance from the center of the membrane; (center) the angle between the transition dipole moment and the *z*-axis; (bottom) angle between the axis perpendicular to the plane and the *z*-axis. These data are taken from a free MD run and are convoluted with Gaussian profile peaks with a full width half maximum of 8°. The errors for the angle distributions are given in Figure S6.

**Fluorescence anisotropy**

When polarized light is applied to a biological environment, the probability of excitation of the probe depends on the angle between the transition state dipole moment and the electric field vector of the incoming electromagnetic radiation. A smaller angle leads to a higher excitation probability. As a consequence, the initial emission after pulsed excitation has a defined polarization. Rotational mobility within a time span determined by the fluorescence lifetime will reduce the fluorescence polarization. The fluorescence anisotropy $r$ is generally defined by means of the fluorescence intensities obtained parallel ($I_{//}$) and perpendicular ($I_\perp$) to the polarization of the excitation light via

$$r = \frac{I_{//} - I_\perp}{I_{//} + 2I_\perp}, \tag{3}$$

when the sample is excited with vertically excited light [84].

For DiI-C18(5) and BNP in lipid bilayers of various composition, the relaxation of $r(t)$ after a δ-pulse excitation was investigated. This relaxation depends on the rotational dynamics, the intrinsic anisotropy $r_0$ (corresponding to the anisotropy at $t$=0) and the conditions of the environment surrounding the light sensitive probe. In agreement with the study by Lipari and Szabo upon the effect of librational motion upon fluorescence depolarization [85] and in line with the theoretical models advocated by Heyn, Jähnig and Ameloot [86–88], the rotational correlation function $C(t)$ is an autocorrelation function and is given in terms of the second order Legendre polynomial $P_2(x)=(3x^2-1)/2$ and the orientation of the transition dipole moment at $t = 0$, $\mu(0)$, and time $t$ after excitation, $\mu(t)$ [34,89]:

$$C(t) = \langle P_2(\mu(0)\mu(t)) \rangle, \tag{4}$$

where the brackets denote the ensemble average, or equivalently, the average over all initial times in the MD calculations, and with $C(t) = \frac{r(t)}{r_0}$ [85]. Since our quantum chemical calculations indicate that the absorption and emission dipoles of the probes under investigation are parallel to each other and as the intrinsic anisotropy $r_0$ or the anisotropy at time $t = 0$ for 1-photon excitation depends on the angle $\delta$ between both dipoles via [84]:

$$r_0 = \frac{2}{5} P_2(\cos \delta), \tag{5}$$

a maximum value of $r_0 = 0.4$ has been considered.

Being embedded in a lipid bilayer, the fluorophore has a limited rotational freedom. The fluorescence lifetime (ranging from hundreds of picoseconds to a few nanoseconds) sets a time window over which the rotational motions can be monitored in an experimental context. In line with previous theoretical and experimental analysis [34,35], a double exponential function is used to describe the rotational correlation function:

$$C(t) = \beta_1 \exp(-t/\theta_1) + \beta_2 \exp(-t/\theta_2) + C_\infty, \tag{6}$$

where $\theta_1$ and $\theta_2$ are correlation times. The $C_\infty$ constant reflects that the rotational correlation function, and therefore the fluorescence anisotropy, does not decay to zero. One can define the mean correlation time $\langle \theta \rangle$ as: $\langle \theta \rangle = \dfrac{\sum_i \beta_i \theta_i^2}{\sum_i \beta_i \theta_i}$.

The results of the analysis are given in Table 1. The quality of the fit was tested by the $\chi^2$ analysis. As our fit leads here to a deviation in the order of barely $10^{-6}$, the high quality of the function used is ensured with a time window up to 25 ns. The $C_\infty$ parameter in the $S_o$ phase for both DiI-C18(5) and BNP are the highest ones in the range of investigated environments, pointing at a particularly confined freedom of rotation. The residual $C_\infty$ for both compounds decreases when a more fluid-like lipid environment is considered. It can also be seen that the $L_d$ phase of DPPC displays a slightly smaller constant than the one of DOPC in the same phase. From our analysis, it has been found that $C_\infty$ ($S_o$) > $C_\infty$ (DOPC, $L_d$) > $C_\infty$ (DPPC at 323K, $L_d$) > $C_\infty$ ($L_o$). These inequalities have to be put in relation to the nature and packing of the various membranes. For the difference between the results for the $L_d$ and $L_o$ phase, the particular position of DiI-C18(5) in the SM:Chol membrane and the presence of the free volumes with water can be recalled. The restricted motions of the probes are finally confirmed by the smaller (larger) relaxation time constants $\theta_1$ ($\theta_2$). For DPPC ($L_d$) and DiI-C18(3), Gullapalli *et al.* reported $\theta_1 = 0.99$ ns and $\theta_2 = 6.9$ ns for the fast and slow components [34]. These values have to be compared with the ones of 0.11 ns ($\theta_1$) and 11.57 ns ($\theta_2$) found for DiI-C18(5) in this study. The values reported by Ariola *et al.*, who studied DiI-C12(3) in the DOPC ($L_d$) membrane, can be compared with the ones of Gullapalli *et al.* and amount to $\theta_1 = 1.2$ ns and $\theta_2 = 9.6$ ns [90]. The obtained time constants for the SM:Chol membrane with not only a very low fast component but also a low slow component point at the special place of the DiI-C18(5) probe: a low steric hindrance of the chromophore is seen in the neighborhood of the top of the lipid acyl chains, while also the collective motion of the lipids in the membrane does not stretch the decay of the rotational autocorrelation function.

A steady-state fluorescence anisotropy of ~0.35 has been measured for BNP in the DPPC $S_o$ phase, while it decreased to ~0.15 upon transition to the $L_d$ phase. The fluorescence lifetimes of this probe reaching up to 4.4 ± 0.2 ns were found to be independent of the phase and the temperature of the lipid system [35]. Especially for BNP, changes in fluorescence anisotropy can consequently be entirely ascribed to restricted tumbling motions of the probe, which are described by Table 1 with the two relaxation times and the limiting anisotropy at long times. From the time constants, it can be seen that the mean relaxation times are larger for BNP than for DiI-C18(5). As the carbocyanines are known to have a shorter fluorescence lifetime of ~1.0 ns [26], the steady state fluorescence anisotropy of BNP is thereupon more sensitive to slower rotational motions than DiI-C18(5). The presented data confirm therefore successfully the assumptions made for BNP at the time of its synthesis [35].

The profoundly low value of 0.12 for $C_\infty$ in SM:Chol as well as the small associated average decay time of 0.43 ns found for DiI-C18(5) point at a strongly pronounced decay of the fluorescence anisotropy and might be another manifestation of the presence of free volumes and a high amount of water molecules in the top polar region of the lipid bilayer. As depicted in Figure 4, the tails of the probe are located along with the acyl tails of the lipids in the membrane. The tails of DiI in SM:Chol are almost parallel to the *z*-axis as can be deduced from the angle of ~170° between the *z*-axis and the vector described by the first and one of the last carbon atoms of the acyl tails of DiI (See Figure S3). Differences between the fitted parameters (*e.g.* $C_\infty$ ~ 0.62 and 0.41 for DiI-C18(5) and BNP in DPPC($L_d$) – or 0.12 for DiI-C18(5) and 0.69 for BNP) for DiI-C18(5) and BNP can finally be related to the differences in position of the probes in the lipid bilayer. It is again an indication for the fundamental differences between the two probes. The anisotropy results, together with the Gibbs energy profiles of DiI-C18(5) embedded in the various lipid bilayers, correct and supplement the image for DiI-C18(5) provided in [76] as the probe is not found to perform surface dynamics in the water phase of the membrane but rather tumbles with two relaxation time constants at different distances from the center of the bilayer.

To give an interpretation to the $C_\infty$ parameter, Kinosita *et al.* proposed in 1977 a so-called 'wobbling in a cone' model, in which the transition dipole and the symmetry axis of the probe are assumed to move without restriction in a cone fixed with respect to the membrane [91]. The model relates the $C_\infty$ parameter to half the cone angle such that a large value of $C_\infty$ corresponds to a small cone angle. It can be remarked that the transition state dipole moments for DiI-C18(5) and for BNP are not oriented along the lipid tails of the respective membranes, which invalidates the 'wobbling in a cone' model [85].

When DiI-C18(5) is approximated to a rod which is oriented along the backbone of the probe, Kinosita's other model of 'wobbling outside the cone' could be considered [91], which describes a spatial angle which is avoided by the transition state dipole moment. The analysis of the spherical coordinates (See Figure S7) gives a limited range for the angle between the transition dipole moment and the z-axis, which would be natural for any model describing a wobbling motion, *as well as* for the movement in the plane of the membrane described by the angle $\varphi$. It is this hindrance in $\varphi$ which invalidates the 'wobbling outside the cone' model as it assumes a free movement of the emission dipole moment for this angle. In the figure, it is also seen that the restriction of the motion of DiI-C18(5) in the plane is less severe for the $L_d$ phases than it is for the $S_o$ and $L_o$ phase. These plots are disentangled in Figures S8 and S9, in which the densities for the individual movements along the $\varphi$ and $\theta$ angles are given. All in all, for DOPC($L_d$) and DPPC ($L_d$), the probe can move in the plane of the membrane over angles of 1.4 and 1.2 radians (~80° and ~70°), respectively. For DPPC ($S_o$) and SM:Chol ($L_o$), the range of $\varphi$ amounts to 0.3 and 0.4 radians (~17° and ~22°), respectively. Discarding small artefacts due to a limited simulation time, these plots are found to be symmetric around 0° for $\varphi$ and 90° for $\theta$. For DiI-C18(5) embedded in SM:Chol, the theta angle is however exclusively restricted to the first quadrant.

Since the tails of the DiI-C18(5) probe can be compared to e.g. the two acyl chains of a DPPC lipid and making abstract of the flexibility of the upper bonds and the out-of-plane distortions of the upper dihedral angles in the tails, the tumbling motion of the backbone and therefore transition state dipole moment of DiI-C18(5) can be related to any wobbling motion of the neighboring lipids. The 3-dimensional movement of the transition state dipole moment is given in Figure 8, showing the specific and restricted movement of the dye up to a timescale of 100 ps. For DPPC ($L_d$), the movement of the probe can be read and a connection can be made with the areas of high density in the plane of the molecule, as visualized by the angle φ in Figure S8. The transition dipole moment of the probe describes zones in time with periods of ~60 ns due to a rather constrained movement in phase with the neighboring lipids and exhibits herein a motion with a smaller solid angle. For DOPC, analogous solid areas are seen. For SM:Chol ($L_o$), the zones are described in ~75 ns, while for DPPC ($S_o$), this period increases to almost 90 ns.

Table 1 – Pre-exponential parameters $\beta$ and rotational correlation time $\theta$ for DiI-C18(5) and BNP in the four considered environments. All rotational correlation times are given in *ns*.[a]

|  |  | $\beta_1$ | $\theta_1$ | $\beta_2$ | $\theta_2$ | $C_\infty$ | $\langle\theta\rangle$ |
|---|---|---|---|---|---|---|---|
| **DiI-C18(5)** | DOPC ($L_d$) | 0.02 | 0.07 | 0.09 | 2.93 | 0.89 | 2.91 |
|  | DPPC ($S_o$) | 0.02 | 0.05 | 0.02 | 2.67 | 0.97 | 2.63 |
|  | DPPC ($L_d$) | 0.04 | 0.11 | 0.34 | 11.57 | 0.62 | 11.55 |
|  | SM:Chol ($L_o$) | 0.47 | 0.06 | 0.41 | 0.48 | 0.12 | 0.43 |
| **BNP** | DOPC ($L_d$) | 0.06 | 0.39 | 0.27 | 24.69 | 0.65 | 24.60 |
|  | DPPC ($S_o$) | 0.04 | 0.08 | 0.03 | 7.97 | 0.93 | 7.83 |
|  | DPPC ($L_d$) | 0.11 | 0.49 | 0.45 | 19.31 | 0.41 | 19.20 |
|  | SM:Chol ($L_o$) | 0.06 | 0.05 | 0.25 | 15.57 | 0.69 | 15.56 |

[a] The mean correlation time $\langle\theta\rangle$ and the $C_\infty$ are also reported.

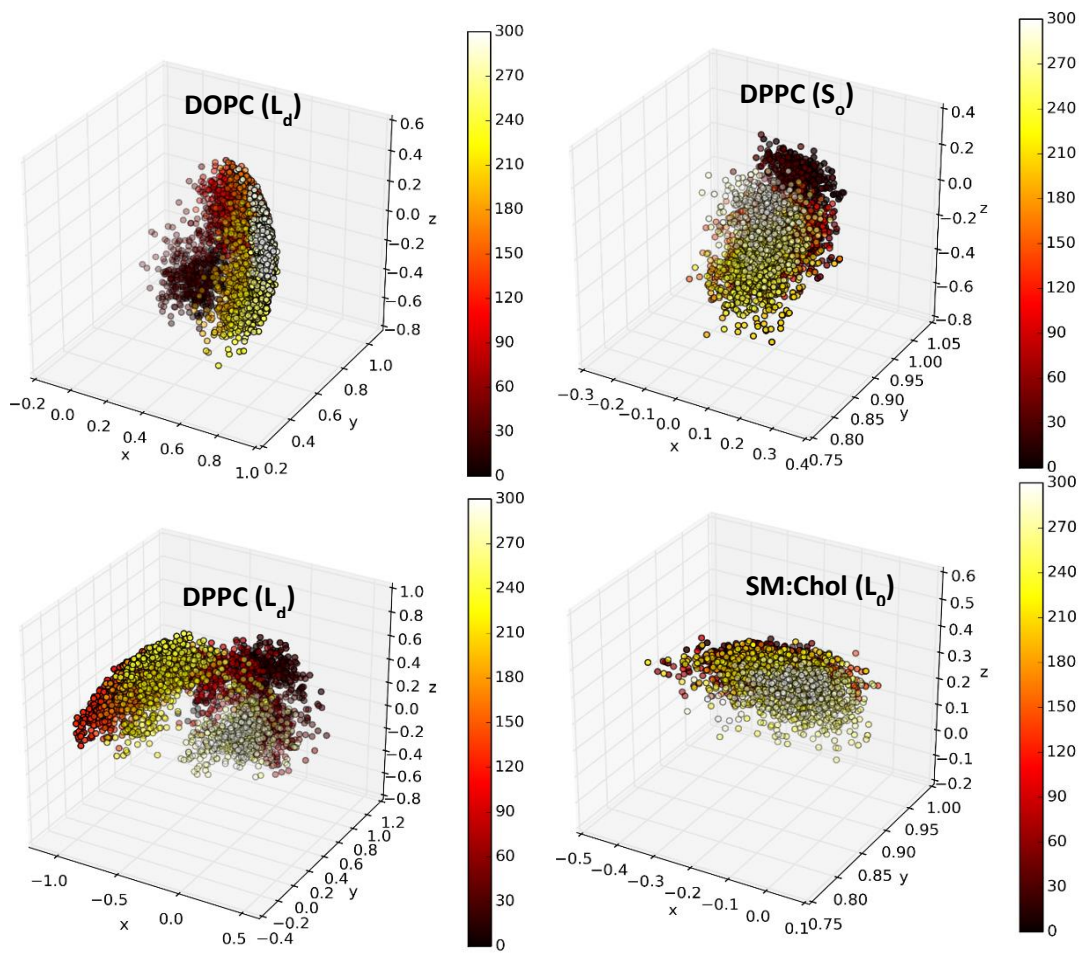

**Figure 8:** The movement of the transition state dipole moment vector of DiI-C18(5) along the MD trajectory. All vectors have been translated to the origin. One dot corresponds to 100 ps; the time runs from 0 ns (black) to 300 ns (white), as indicated by the color bar.

**Conclusions and outlook**

The behavior of BNP and DiI-C18(5) molecular probes was investigated in various lipid bilayers in three different phases. By means of demanding MD simulations, the Gibbs free energy profiles of both probes showed that they preferentially partition into the $S_o$ phase of the DPPC bilayer rather than in the $L_d$ phase of the DOPC bilayer. The $L_o$ phase of a 2:1 SM:Chol mixture was also preferred with respect to the $L_d$ phase.

The positions and orientations of the probes are primordial to anticipate their optical properties *in situ*, *e.g.*, in biological membranes. The depths of insertion differ depending on the phase, and that relative to this, the probes in the SM:Chol mixture are stabilized more towards the polar head group region of the membrane. The orientation of the transition dipole moment is very different with the two probes: for DiI-C18(5), the angle between the transition dipole moment and the $z$-axis in DOPC ($L_d$) is closer to a perfect 90° value than for the rather new probe BNP. A striking difference is however seen for the molecules in the DPPC ($L_d$) phase, for which the distribution of the angle ranges from 70° to 80° for DiI-C18(5), while for BNP it peaks at around 85°. From investigations of the membrane density and supported by simulations of the fluorescence anisotropy, it follows that in the SM:Chol ($L_o$) phase, a high amount of water molecules is found in the vicinity of the probes and that the embedded probes are less restricted in their movement than when they are surrounded by the other membrane phases.

Although the blue fluorescing BNP probe has been introduced as an alternative for the older yellow DiI-C18(5) one, it has been proven that they may behave differently with respect to their interaction with membranes. It is expected that the differences in position and orientation in various biological membranes will affect the linear and more the non-linear absorption spectra. The current research opens therefore a gateway towards a better investigation of the properties of biological membranes and tissues using nonlinear and fluorescent properties of selective molecular probes.

**Supplementary information**

Area per lipid along the simulated trajectory for the various membranes; order-parameters for the various membrane phases for the sn-1 and sn-2 tails; illustrations of the DiI-C18(5) and BNP probes in the different environments under investigation in the current study; radial distribution functions of DiI-C18(5) and surrounding water molecules for the considered membranes; angle with the $z$-axis of the vector described by the first and fifteenth carbon atom of the acyl tails of DiI; distribution of the vector of the transition dipole moment in spherical coordinates $\theta$ and $\varphi$; density plots for the vector of the transition dipole moment in function of the azimuthal angle $\varphi$; density plots for the vector of the transition dipole moment in function of the angle $\theta$; .itp-files for DiI-C18(5) and BNP.

**Acknowledgments**

The authors thank the PDC Center for High Performance Computing (local CAC project 1401-002, *Lindgren* cluster), the Flemish Supercomputer Centre (VSC) (Flanders, Belgium) and the Herculesstichting (*muk tier-1* cluster) (Flanders, Belgium), as well as the "Consortium des Équipements de Calcul Intensif" (Céci, with their *Lemaitre2*, *Nic4*, *Hmem*, *Hercules* and *Dragon1* clusters) of the Association Wallonie-Bruxelles for their generous support in terms of computational time. The authors also acknowledge the Swedish SNAC consortium for the medium projects SNIC 2016/1-87 and SNIC 1-415, as well as the small one SNIC 2015/4-44. PT, GF and FDM thank CALI. PT thanks the Czech Science Foundation (P208/12/G016) and National Program of Sustainability I from the Ministry-of-Youth, Education and Sports of the Czech Republic (LO1305). S.O. is grateful to the Center for Quantum Materials and Nordita for his funding in Sweden. S.O. acknowledges the National Science Centre, Poland, grant UMO-2015/19/P/ST4/03636 for the funding from the European Union's Horizon 2020 research and innovation program under the Marie Skłodowska-Curie grant agreement No. 665778. M. P. thanks the Department of Theoretical Chemistry and Biology for her post-doctoral funding at KTH. S. K. is grateful to the Fonds National de la Recherche Scientifique, the French speaking branch of the Belgian National Science foundation, for his postdoctoral funding ('Chargé de Recherches') in Liège. S. K. would also like to thank the group in Limoges for the offered hospitality during his various long term research stays.

**References**

1 M. A. Lindsay, Target discovery, *Nat. Rev. Drug Discov.*, 2003, **2**, 831–838.
2 I. Green, R. Christison, C. J. Voyce, K. R. Bundell and M. A. Lindsay, Protein transduction domains: are they delivering?, *Trends Pharmacol. Sci.*, 2003, **24**, 213–215.
3 U. Haberkorn, W. Mier, K. Kopka, C. Herold-Mende, A. Altmann and J. Babich, Identification of Ligands and Translation to Clinical Applications, *J. Nucl. Med.*, 2017, **58**, 27S-33S.
4 A. Weiss and P. Nowak-Sliwinska, Current Trends in Multidrug Optimization: An Alley of Future Successful Treatment of Complex Disorders, *SLAS Technol. Transl. Life Sci. Innov.*, 2017, **22**, 254–275.
5 I. Nakazawa and M. Iwaizumi, A Role of the Cancer Cell-Membrane Fluidity in the Cancer Metastases -, *Tohoku J. Exp. Med.*, 1989, **157**, 193–198.
6 D. Voet and J. G. Voet, *Biochemistry*, John Wiley and Sons, Hoboken, NJ, 2011.
7 A. Gorin, L. Gabitova and I. Astsaturov, Regulation of cholesterol biosynthesis and cancer signaling, *Curr. Opin. Pharmacol.*, 2012, **12**, 710–716.
8 A. Fages, T. Duarte-Salles, M. Stepien, P. Ferrari, V. Fedirko, C. Pontoizeau, A. Trichopoulou, K. Aleksandrova, A. Tjonneland, A. Olsen, F. Clavel-Chapelon, M.-C. Boutron-Ruault, G. Severi, R. Kaaks, T. Kuhn, A. Floegel, H. Boeing, P. Lagiou, C. Bamia, D. Trichopoulos, D. Palli, V. Pala, S. Panico, R. Tumino, P. Vineis, H. B. Bueno-

de-Mesquita, P. H. Peeters, E. Weiderpass, A. Agudo, E. Molina-Montes, J. Maria Huerta, E. Ardanaz, M. Dorronsoro, K. Sjoberg, B. Ohlsson, K.-T. Khaw, N. Wareham, R. C. Travis, J. A. Schmidt, A. Cross, M. Gunter, E. Riboli, A. Scalbert, I. Romieu, B. Elena-Herrmann and M. Jenab, Metabolomic profiles of hepatocellular carcinoma in a European prospective cohort, *Bmc Med.*, 2015, **13**, 242.

9  K. Bartel, M. Winzi, M. Ulrich, A. Koeberle, D. Menche, O. Werz, R. Mueller, J. Guck, A. M. Vollmar and K. von Schwarzenberg, V-ATPase inhibition increases cancer cell stiffness and blocks membrane related Ras signaling - a new option for HCC therapy, *Oncotarget*, 2017, **8**, 9476–9487.

10 G. van Meer, D. R. Voelker and G. W. Feigenson, Membrane lipids: where they are and how they behave, *Nat. Rev. Mol. Cell Biol.*, 2008, **9**, 112–124.

11 D. Recktenwald and H. Mcconnell, Phase-Equilibria in Binary-Mixtures of Phosphatidylcholine and Cholesterol, *Biochemistry (Mosc.)*, 1981, **20**, 4505–4510.

12 K. Simons and W. L. C. Vaz, Model systems, lipid rafts, and cell membranes, *Annu. Rev. Biophys. Biomol. Struct.*, 2004, **33**, 269–295.

13 T. P. W. McMullen, R. N. A. H. Lewis and R. N. McElhaney, Cholesterol–phospholipid interactions, the liquid-ordered phase and lipid rafts in model and biological membranes, *Curr. Opin. Colloid Interface Sci.*, 2004, **8**, 459–468.

14 H. Sprong, P. van der Sluijs and G. van Meer, How proteins move lipids and lipids move proteins, *Nat. Rev. Mol. Cell Biol.*, 2001, **2**, 504–513.

15 K. Simons and J. L. Sampaio, Membrane Organization and Lipid Rafts, *Cold Spring Harb. Perspect. Biol.*, 2011, **3**, a004697–a004697.

16 L. K. Buehler, *Cell membranes*, Garland Science, Abingdon, UK, 2016.

17 T. Baumgart, G. Hunt, E. R. Farkas, W. W. Webb and G. W. Feigenson, Fluorescence probe partitioning between Lo/Ld phases in lipid membranes, *Biochim. Biophys. Acta BBA - Biomembr.*, 2007, **1768**, 2182–2194.

18 N. Kahya, D. Scherfeld, K. Bacia, B. Poolman and P. Schwille, Probing Lipid Mobility of Raft-exhibiting Model Membranes by Fluorescence Correlation Spectroscopy, *J. Biol. Chem.*, 2003, **278**, 28109–28115.

19 J. Juhasz, J. H. Davis and F. J. Sharom, Fluorescent probe partitioning in giant unilamellar vesicles of "lipid raft" mixtures, *Biochem. J.*, 2010, **430**, 415–423.

20 K. Bacia, D. Scherfeld, N. Kahya and P. Schwille, Fluorescence Correlation Spectroscopy Relates Rafts in Model and Native Membranes, *Biophys. J.*, 2004, **87**, 1034–1043.

21 D. Scherfeld, N. Kahya and P. Schwille, Lipid dynamics and domain formation in model membranes composed of ternary mixtures of unsaturated and saturated phosphatidylcholines and cholesterol, *Biophys. J.*, 2003, **85**, 3758–3768.

22 R. F. M. de Almeida, A. Fedorov and M. Prieto, Sphingomyelin/phosphatidylcholine/cholesterol phase diagram: Boundaries and composition of lipid rafts, *Biophys. J.*, 2003, **85**, 2406–2416.

23 P. Uppamoochikkal, S. Tristram-Nagle and J. F. Nagle, Orientation of Tie-Lines in the Phase Diagram of DOPC/DPPC/Cholesterol Model Biomembranes, *Langmuir*, 2010, **26**, 17363–17368.

24 N. Bezlyepkina, R. S. Gracia, P. Shchelokovskyy, R. Lipowsky and R. Dimova, Phase Diagram and Tie-Line Determination for the Ternary Mixture DOPC/eSM/Cholesterol, *Biophys. J.*, 2013, **104**, 1456–1464.

25 L. M. Loew, *Spectroscopic Membrane Probes 1*, CRC Press, Boca Raton, Florida, 1988.

26 B. Packard and D. Wolf, Fluorescence Lifetimes of Carbocyanine Lipid Analogs in Phospholipid-Bilayers, *Biochemistry (Mosc.)*, 1985, **24**, 5176–5181.

27 P. J. Sims, A. S. Waggoner, C.-H. Wang and J. F. Hoffman, Mechanism by which cyanine dyes measure membrane potential in red blood cells and phosphatidylcholine vesicles, *Biochemistry (Mosc.)*, 1974, **13**, 3315–3330.

28 B. C. Stevens and T. Ha, Discrete and heterogeneous rotational dynamics of single membrane probe dyes in gel phase supported lipid bilayer, *J. Chem. Phys.*, 2004, **120**, 3030–3039.

29 M. Perin and R. Macdonald, Interactions of Liposomes with Planar Bilayer-Membranes, *J. Membr. Biol.*, 1989, **109**, 221–232.

30 S. Draxler, M. Lippitsch and F. Aussenegg, Long-Range Excitation-Energy Transfer in Langmuir Blodgett Multilayer Systems, *Chem. Phys. Lett.*, 1989, **159**, 231–234.

31 C. Spink, M. Yeager and G. Feigenson, Partitioning Behavior of Indocarbocyanine Probes Between Coexisting Gel and Fluid Phases in Model Membranes, *Biochim. Biophys. Acta*, 1990, **1023**, 25–33.

32 D. Wolf, Determination of the Sidedness of Carbocyanine Dye Labeling of Membranes, *Biochemistry (Mosc.)*, 1985, **24**, 582–586.

33 A. M. Davey, R. P. Walvick, Y. Liu, A. A. Heikal and E. D. Sheets, Membrane order and molecular dynamics associated with IgE receptor cross-linking in mast cells, *Biophys. J.*, 2007, **92**, 343–355.

34 R. R. Gullapalli, M. C. Demirel and P. J. Butler, Molecular dynamics simulations of DiI-C18(3) in a DPPC lipid bilayer, *Phys. Chem. Chem. Phys.*, 2008, **10**, 3548.

35 M. Bacalum, L. Wang, S. Boodts, P. Yuan, V. Leen, N. Smisdom, E. Fron, S. Knippenberg, G. Fabre, P. Trouillas, D. Beljonne, W. Dehaen, N. Boens and M. Ameloot, A Blue-Light-Emitting BODIPY Probe for Lipid Membranes, *Langmuir*, 2016, **32**, 3495–3505.

36 Li and J.-X. Cheng, Coexisting Stripe- and Patch-Shaped Domains in Giant Unilamellar Vesicles †, *Biochemistry (Mosc.)*, 2006, **45**, 11819–11826.

37 D. Bassolinoklimas, H. Alper and T. Stouch, Solute Diffusion in Lipid Bilayer-Membranes - an Atomic-Level Study by Molecular-Dynamics Simulation, *Biochemistry (Mosc.)*, 1993, **32**, 12624–12637.

38 D. Bassolinoklimas, H. Alper and T. Stouch, Mechanism of Solute Diffusion Through Lipid Bilayer-Membranes by Molecular-Dynamics Simulation, *J. Am. Chem. Soc.*, 1995, **117**, 4118–4129.

39 S. J. Marrink and H. J. C. Berendsen, Permeation process of small molecules across lipid membranes studied by molecular dynamics simulations, *J. Phys. Chem.*, 1996, **100**, 16729–16738.

40 R. O. Dror, R. M. Dirks, J. P. Grossman, H. Xu and D. E. Shaw, ed. D. C. Rees, Annual Reviews, Palo Alto, 2012, vol. 41, pp. 429–452.

41 M. Karplus and J. A. McCammon, Molecular dynamics simulations of biomolecules, *Nat. Struct. Biol.*, 2002, **9**, 646–652.

42 M. Karplus and J. Kuriyan, Molecular dynamics and protein function, *Proc. Natl. Acad. Sci. U. S. A.*, 2005, **102**, 6679–6685.

43 T.-X. Xiang and B. D. Anderson, Liposomal drug transport: A molecular perspective from molecular dynamics simulations in lipid bilayers, *Adv. Drug Deliv. Rev.*, 2006, **58**, 1357–1378.

44 D. Bemporad, C. Luttmann and J. W. Essex, Behaviour of small solutes and large drugs in a lipid bilayer from computer simulations, *Biochim. Biophys. Acta-Biomembr.*, 2005, **1718**, 1–21.

45 M. B. Boggara and R. Krishnamoorti, Partitioning of Nonsteroidal Antiinflammatory Drugs in Lipid Membranes: A Molecular Dynamics Simulation Study, *Biophys. J.*, 2010, **98**, 586–595.

46 H. S. Muddana, R. R. Gullapalli, E. Manias and P. J. Butler, Atomistic simulation of lipid and DiI dynamics in membrane bilayers under tension, *Phys. Chem. Chem. Phys.*, 2011, **13**, 1368–1378.

47 M. Paloncyova, K. Berka and M. Otyepka, Convergence of Free Energy Profile of Coumarin in Lipid Bilayer, *J. Chem. Theory Comput.*, 2012, **8**, 1200–1211.

48 N. A. Murugan, R. Apostolov, Z. Rinkevicius, J. Kongsted, E. Lindahl and H. Agren, Association Dynamics and Linear and Nonlinear Optical Properties of an N-Acetylaladanamide Probe in a POPC Membrane, *J. Am. Chem. Soc.*, 2013, **135**, 13590–13597.

49 B. Mennucci, M. Caricato, F. Ingrosso, C. Cappelli, R. Cammi, J. Tomasi, G. Scalmani and M. J. Frisch, How the environment controls absorption and fluorescence spectra of PRODAN: A quantum-mechanical study in homogeneous and heterogeneous media, *J. Phys. Chem. B*, 2008, **112**, 414–423.

50 G. Parisio, A. Marini, A. Biancardi, A. Ferrarini and B. Mennucci, Polarity-Sensitive Fluorescent Probes in Lipid Bilayers: Bridging Spectroscopic Behavior and Microenvironment Properties, *J. Phys. Chem. B*, 2011, **115**, 9980–9989.

51 S. Osella, N. A. Murugan, N. K. Jena and S. Knippenberg, Investigation into Biological Environments through (Non)linear Optics: A Multiscale Study of Laurdan Derivatives, *J. Chem. Theory Comput.*, 2016, **12**, 6169–6181.

52 S. Osella and S. Knippenberg, Triggering On/Off States of Photoswitchable Probes in Biological Environments, *J. Am. Chem. Soc.*, 2017, 4418–4428.

53 D. Van Der Spoel, E. Lindahl, B. Hess, G. Groenhof, A. E. Mark and H. J. C. Berendsen, GROMACS: fast, flexible, and free, *J. Comput. Chem.*, 2005, **26**, 1701–1718.

54 B. Hess, C. Kutzner, D. van der Spoel and E. Lindahl, GROMACS 4: Algorithms for Highly Efficient, Load-Balanced, and Scalable Molecular Simulation, *J. Chem. Theory Comput.*, 2008, **4**, 435–447.

55 S. A. Pandit, S. Vasudevan, S. W. Chiu, R. Jay Mashl, E. Jakobsson and H. L. Scott, Sphingomyelin-Cholesterol Domains in Phospholipid Membranes: Atomistic Simulation, *Biophys. J.*, 2004, **87**, 1092–1100.

56 S. A. Pandit, S.-W. Chiu, E. Jakobsson, A. Grama and H. L. Scott, Cholesterol packing around lipids with saturated and unsaturated chains: a simulation study, *Langmuir ACS J. Surf. Colloids*, 2008, **24**, 6858–6865.

57 S. W. Chiu, S. Vasudevan, E. Jakobsson, R. J. Mashl and H. L. Scott, Structure of sphingomyelin bilayers: a simulation study, *Biophys. J.*, 2003, **85**, 3624–3635.

58 S. A. Pandit, S.-W. Chiu, E. Jakobsson, A. Grama and H. L. Scott, Cholesterol surrogates: a comparison of cholesterol and 16:0 ceramide in POPC bilayers, *Biophys. J.*, 2007, **92**, 920–927.

59 T. Darden, D. York and L. Pedersen, Particle mesh Ewald: An N·log(N) method for Ewald sums in large systems, *J. Chem. Phys.*, 1993, **98**, 10089–10092.

60 B. Hess, H. Bekker, H. J. C. Berendsen and J. Fraaije, LINCS: A linear constraint solver for molecular simulations, *J. Comput. Chem.*, 1997, **18**, 1463–1472.

61 S. Nose, A Unified Formulation of the Constant Temperature Molecular-Dynamics Methods, *J. Chem. Phys.*, 1984, **81**, 511–519.

62 W. G. Hoover, Canonical dynamics: Equilibrium phase-space distributions, *Phys. Rev. A*, 1985, **31**, 1695–1697.

63 M. Parrinello and A. Rahman, Polymorphic Transitions in Single-Crystals - a New Molecular-Dynamics Method, *J. Appl. Phys.*, 1981, **52**, 7182–7190.

64 A. W. Schuttelkopf and D. M. F. van Aalten, PRODRG: a tool for high-throughput crystallography of protein-ligand complexes, *Acta Crystallogr. Sect. -Biol. Crystallogr.*, 2004, **60**, 1355–1363.

65 J. M. Wang, P. Cieplak and P. A. Kollman, How well does a restrained electrostatic potential (RESP) model perform in calculating conformational energies of organic and biological molecules?, *J. Comput. Chem.*, 2000, **21**, 1049–1074.

66 A. D. Becke, Density-functional thermochemistry. III. The role of exact exchange, *J. Chem. Phys.*, 1993, **98**, 5648–5652.

67 C. Lee, W. Yang and R. G. Parr, Development of the Colle-Salvetti correlation-energy formula into a functional of the electron density, *Phys. Rev. B*, 1988, **37**, 785–789.

68 T. Dunning, Gaussian-Basis Sets for Use in Correlated Molecular Calculations .1. the Atoms Boron Through Neon and Hydrogen, *J. Chem. Phys.*, 1989, **90**, 1007–1023.

69 Y. Duan, C. Wu, S. Chowdhury, M. C. Lee, G. M. Xiong, W. Zhang, R. Yang, P. Cieplak, R. Luo, T. Lee, J. Caldwell, J. M. Wang and P. Kollman, A point-charge force field for molecular mechanics simulations of proteins based on condensed-phase quantum mechanical calculations, *J. Comput. Chem.*, 2003, **24**, 1999–2012.

70 E. D. Pietro, G. Cardini and V. Schettino, Ab initio molecular dynamics study of the hydrolysis reaction of diborane, *Phys. Chem. Chem. Phys.*, 2007, **9**, 3857–3863.

71 M. Paloncýová, K. Berka and M. Otyepka, Convergence of free energy profile of coumarin in lipid bilayer, *J. Chem. Theory Comput.*, 2012, **8**, 1200–1211.

72 M. Paloncýová, G. Fabre, R. H. DeVane, P. Trouillas, K. Berka and M. Otyepka, Benchmarking of Force Fields for Molecule–Membrane Interactions, *J. Chem. Theory Comput.*, 2014, **10**, 4143–4151.

73 M. Paloncýová, R. DeVane, B. Murch, K. Berka and M. Otyepka, Amphiphilic Drug-Like Molecules Accumulate in a Membrane below the Head Group Region, *J. Phys. Chem. B*, 2014, **118**, 1030–1039.

74 M. G. Wolf, M. Hoefling, C. Aponte-Santamaría, H. Grubmüller and G. Groenhof, g_membed: Efficient insertion of a membrane protein into an equilibrated lipid bilayer with minimal perturbation, *J. Comput. Chem.*, 2010, **31**, 2169–2174.

75 S. Buchoux, FATSLiM: a fast and robust software to analyze MD simulations of membranes, *Bioinformatics*, 2017, **33**, 133–134.

76 M. M. G. Krishna, A. Srivastava and N. Periasamy, Rotational dynamics of surface probes in lipid vesicles, *Biophys. Chem.*, 2001, **90**, 123–133.

77 S. Lopes and M. Castanho, Does aliphatic chain length influence carbocyanines' orientation in supported lipid multilayers?, *J. Fluoresc.*, 2004, **14**, 281–287.

78 P. R. Maulik and G. G. Shipley, Interactions of N-stearoyl sphingomyelin with cholesterol and dipalmitoyl phosphatidylcholine in bilayer membranes, *Biophys. J.*, 1996, **70**, 2256–2265.

79 D. Poger and A. E. Mark, On the Validation of Molecular Dynamics Simulations of Saturated and cis-Monounsaturated Phosphatidylcholine Lipid Bilayers: A Comparison with Experiment, *J. Chem. Theory Comput.*, 2010, **6**, 325–336.

80 C. L. Wennberg, D. van der Spoel and J. S. Hub, Large Influence of Cholesterol on Solute Partitioning into Lipid Membranes, *J. Am. Chem. Soc.*, 2012, **134**, 5351–5361.

81 F. Di Meo, G. Fabre, K. Berka, T. Ossman, B. Chantemargue, M. Paloncyova, P. Marquet, M. Otyepka and P. Trouillas, In silico pharmacology: Drug membrane partitioning and crossing, *Pharmacol. Res.*, 2016, **111**, 471–486.

82 L. S. Vermeer, B. L. de Groot, V. Reat, A. Milon and J. Czaplicki, Acyl chain order parameter profiles in phospholipid bilayers: computation from molecular dynamics simulations and comparison with H-2 NMR experiments, *Eur. Biophys. J. Biophys. Lett.*, 2007, **36**, 919–931.

83 D. Axelrod, Carbocyanine Dye Orientation in Red-Cell Membrane Studied by Microscopic Fluorescence Polarization, *Biophys. J.*, 1979, **26**, 557–573.

84 Lakowicz, *Principles of Fluorescence Spectroscopy*, Springer, 3rd edition., 2007.

85 G. Lipari and A. Szabo, Effect of Librational Motion on Fluorescence Depolarization and Nuclear Magnetic-Resonance Relaxation in Macromolecules and Membranes, *Biophys. J.*, 1980, **30**, 489–506.
86 M. Heyn, Determination of Lipid Order Parameters and Rotational Correlation Times from Fluorescence Depolarization Experiments, *Febs Lett.*, 1979, **108**, 359–364.
87 F. Jahnig, Structural Order of Lipids and Proteins in Membranes - Evaluation of Fluorescence Anisotropy Data, *Proc. Natl. Acad. Sci. U. S. A.*, 1979, **76**, 6361–6365.
88 M. Ameloot, H. Hendrickx, W. Herreman, H. Pottel, F. Vancauwelaert and W. Vandermeer, Effect of Orientational Order on the Decay of the Fluorescence Anisotropy in Membrane Suspensions - Experimental-Verification on Unilamellar Vesicles and Lipid Alpha-Lactalbumin Complexes, *Biophys. J.*, 1984, **46**, 525–539.
89 W. van der Meer, H. Pottel, W. Herreman, M. Ameloot, H. Hendrickx and H. Schröder, Effect of orientational order on the decay of the fluorescence anisotropy in membrane suspensions, *Biophys J*, 1984, **46**, 515.
90 F. S. Ariola, D. J. Mudaliar, R. P. Walvick and A. A. Heikal, Dynamics imaging of lipid phases and lipid-marker interactions in model biomembranes, *Phys. Chem. Chem. Phys.*, 2006, **8**, 4517.
91 K. Kinosita Jr, S. Kawato and A. Ikegami, A theory of fluorescence polarization decay in membranes., *Biophys. J.*, 1977, **20**, 289.

**TOC:**

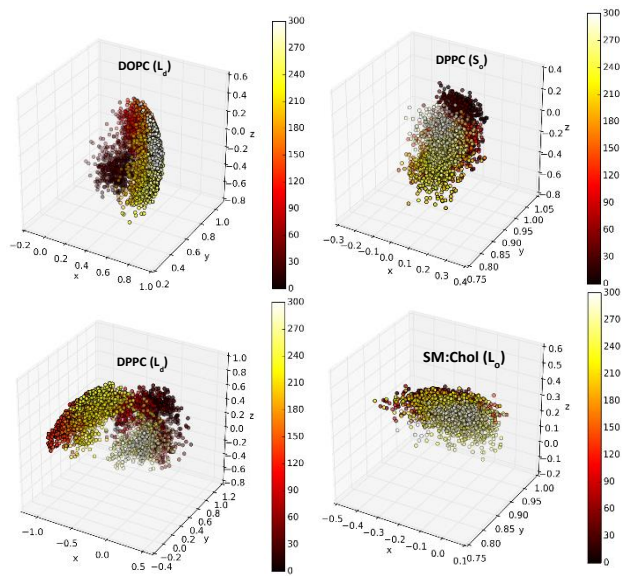

**Movement of DiI-C18(5)**